# Time series features for supporting hydrometeorological explorations and predictions in ungauged locations using large datasets

Georgia Papacharalampous[1,*], and Hristos Tyralis[1,2]

[1] Department of Water Resources and Environmental Engineering, School of Civil Engineering, National Technical University of Athens, Heroon Polytechneiou 5, 15780 Zographou, Greece

[2] Air Force Projects Authority, Hellenic Air Force, Mesogion Avenue 227–231, 15561 Cholargos, Greece

[*] Correspondence: papacharalampous.georgia@gmail.com, tel: +30 69474 98589

**Email addresses and ORCID profiles:** papacharalampous.georgia@gmail.com, gpapacharalampous@hydro.ntua.gr, https://orcid.org/0000-0001-5446-954X (Georgia Papacharalampous); montchrister@gmail.com, hristos@itia.ntua.gr, https://orcid.org/0000-0002-8932-4997 (Hristos Tyralis)

**Abstract:** Regression-based frameworks for streamflow regionalization are built around catchment attributes that traditionally originate from catchment hydrology, flood frequency analysis and their interplay. In this work, we deviated from this traditional path by formulating and extensively investigating the first regression-based streamflow regionalization frameworks that largely emerge from general-purpose time series features for data science and, more precisely, from a large variety of such features. We focused on 28 features that included (partial) autocorrelation, entropy, temporal variation, seasonality, trend, lumpiness, stability, nonlinearity, linearity, spikiness, curvature and others. We estimated these features for daily temperature, precipitation and streamflow time series from 511 catchments, and then merged them within regionalization contexts with traditional topographic, land cover, soil and geologic attributes. Precipitation and temperature features (e.g., the spectral entropy, seasonality strength and lag-1 autocorrelation of the precipitation time series, and the stability and trend strength of the temperature time series) were found to be useful predictors of many streamflow features. The same applies to traditional attributes, such as the catchment mean elevation. Relationships between predictor and dependent variables were also revealed, while the spectral entropy, the seasonality strength and several autocorrelation features of the streamflow time series were found to be more regionalizable than others.

# 1. Introduction

Streamflow regionalization (see its various definitions in He et al. 2011, Table 1) is closely related to the initiative for Predictions in Ungauged Basins (PUB) of the International Association of Hydrological Sciences (IAHS) by Sivapalan et al. (2003). The importance of this initiative is broadly acknowledged in the literature and extensively discussed by other initiatives (e.g., Hrachowitz et al. 2013; Montanari et al. 2013; Blöschl et al. 2019). In summary, the core concept behind streamflow regionalization and PUB is the transfer of information that is useful for streamflow description and modelling from gauged to ungauged sites. This transfer can lead to the reduction of the streamflow modelling uncertainties in ungauged sites and can be facilitated, among others, by regression-based frameworks (see, e.g., the reviews by He et al. 2011; Hrachowitz et al. 2013; Guo et al. 2020), which first establish an empirical relationship between a target and a set of independent variables based on large multi-site datasets, and then utilize the previously established relationship for the information transfer.

The independent variables of the regression may include any catchment attribute that does not rely on flow time series for its estimation, while the target variables of the regression, and thus the information transferred through streamflow regionalization, can take the form of either streamflow time series characterizations or model parameter estimates (see, e.g., the reviews by He et al. 2011; Hrachowitz et al. 2013; Guo et al. 2020), with these latter two attribute categories mostly overlapping in practice, at least from a statistician's point of view. Furthermore, as streamflow modelling through streamflow regionalization mostly appears in the catchment hydrology and the flood frequency analysis fields and in their interplay, the target and independent variables of the regression usually take forms that are relevant to either of these fields. A related literature overview linking streamflow regionalization for flood investigations (see, e.g., the modelling works by Merz and Blöschl 2005; Aziz et al.2014; Rahman et al. 2019; Tyralis et al. 2019b; Rahman et al. 2020; Fischer and Schumann 2021) to streamflow signature regionalization (see, e.g., the modelling works by Beck et al. 2015; Westerberg et al. 2016; Addor et al. 2018; Tyralis et al. 2021b; Laimighofer et al. 2022, as well as the reviews and taxonomies of streamflow signatures by McMillan et al. 2017; McMillan 2020) and catchment model parameter regionalization (see, e.g., the modelling works by Parajka et al. 2005; Oudin et al. 2008; Pool et al. 2021) can be found in Tyralis et al. (2019b).

Notably, streamflow regionalization frameworks that are largely centered around general-purpose time series features (i.e., functions for extracting information from time series data), such as the autocorrelation, partial autocorrelation, entropy, temporal variation, seasonality, trend, lumpiness, stability, nonlinearity, linearity, spikiness and curvature ones appearing in Wang et al. (2006), Fulcher et al. (2013), Fulcher and Jones (2014), Hyndman et al. (2015), Fulcher and Jones (2017), Kang et al. (2017), Fulcher (2018), Kang et al. (2020) and Hyndman et al. (2020), are currently absent from the literature, with the same further applying individually to most of the relevant feature categories. These hold despite the fundamental and practical interest appearing in stochastic (statistical) hydrology for many, if not all, of the above-mentioned feature categories (see, e.g., the central themes, concepts and directions provided by Montanari et al. 2013, and five key features investigated in Papacharalampous and Tyralis 2020), and despite their proven relevance for diverse data science (Donoho 2017) tasks (including hydro-data science tasks, such as those carried out in Papacharalampous et al. 2021, 2022), thereby suggesting a research gap waiting to be filled. The importance of filling this specific gap in general, and of filling it with multiple time series features in particular, becomes even more pronounced if we additionally consider that hydrometeorological conditions should ideally be represented by as many features as possible (Papacharalampous et al. 2021).

Driven by the above considerations, we here propose the estimation of a large variety of general-purpose time series features, including features from all the categories mentioned in the above paragraph and more, from large multi-site datasets comprising temperature, precipitation and streamflow information, and the subsequent transfer of streamflow feature information from gauged to ungauged locations by using the various temperature and precipitation features, together with multiple other catchment attributes, as predictor variables within regression-based streamflow regionalization frameworks. This is possible and largely applicable in practice (similarly to what applies to other forms of regression-based streamflow regionalizations), as a variety of topographic, land cover, soil and geologic attributes might be available for both gauged and ungauged catchments and, at the same time, temperature and precipitation features can be sufficiently estimated for them from widely available remote sending data, in case that earth observations for temperature and precipitation are not available.

Given their scarcity in the previous streamflow regionalization literature, which is also evident from literature reviews on the topic (e.g., He et al. 2011; Hrachowitz et al. 2013; Guo et al. 2020), the various time series feature categories underlying this work, as well as their consideration as a synthesis, constitute new concepts and methodological elements for supporting: (a) regression-based streamflow feature regionalization, and (b) explorations that facilitate an improved understanding of the relationships that can be exploited for this regionalization. Therefore, their usefulness had to be extensively investigated. In this endeavour, we herein designed and performed a series of large-sample experiments, as detailed in Section 2 and as summarized with Figure 1.

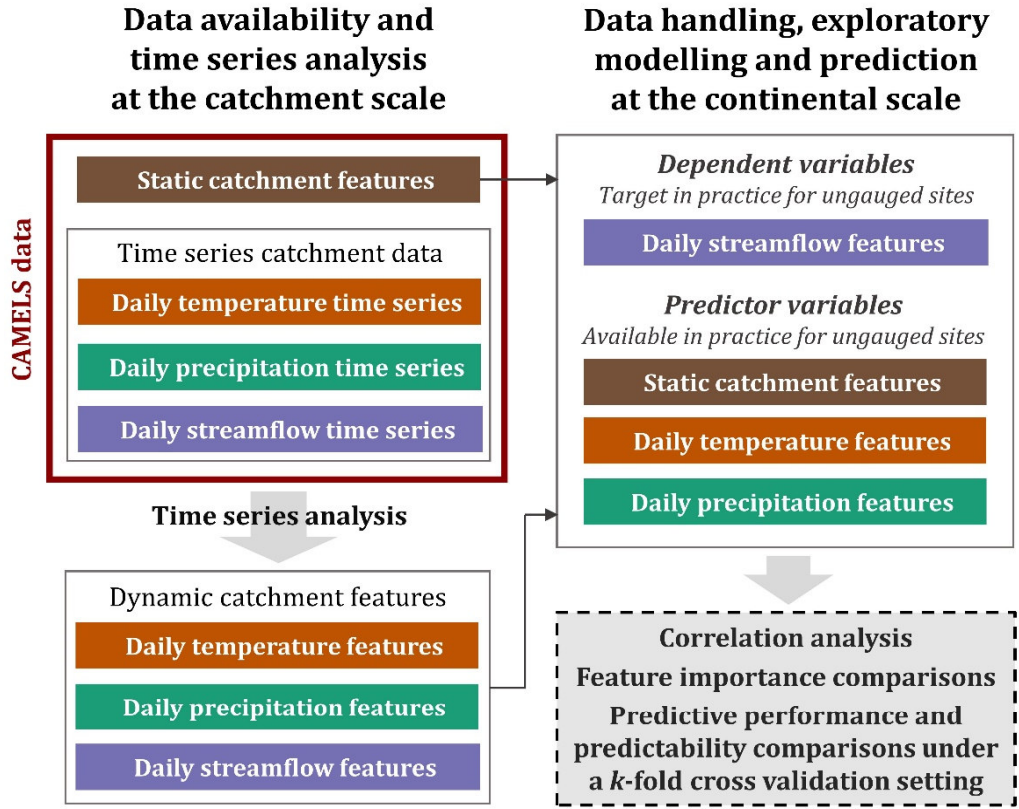

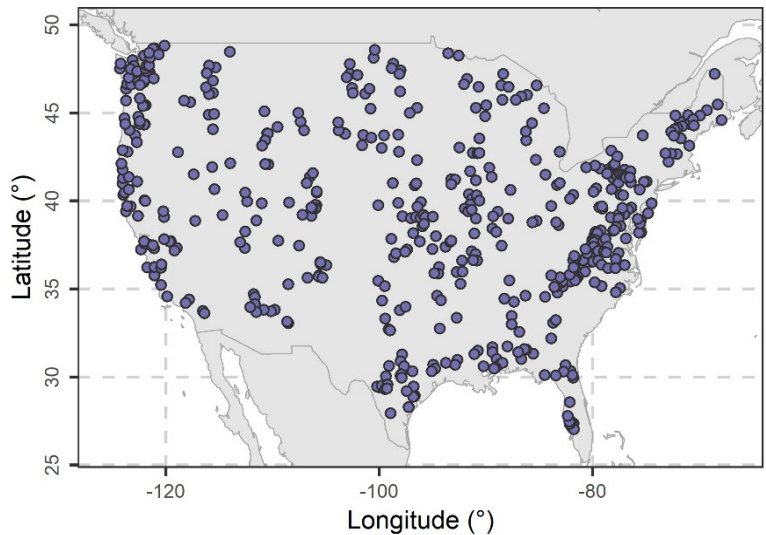

Figure 1. Schematic summarizing the experimental dataset and methodology.

These experiments relied on a well-established hydrological dataset (see Section 2.1) and were implemented using open statistical software according to the details provided in 0. They were conducted at the daily time scale; nonetheless, their underlying methodological framework is also applicable to other (including mixed) time scales after small adaptations in the time series analysis for feature estimation. This analysis was herein performed according to Section 2.2, while a correlation analysis, comparisons of the various potential predictors (which included catchment attributes that were available in the experimental dataset, and daily temperature and precipitation features; see Sections 2.1 and 2.2) with respect to their importance in regionalizing daily streamflow features, predictive performance comparisons and a comparison of the daily streamflow features with respect to their predictability were performed according to Sections 2.3, 2.4, 2.5 and 2.6, respectively. The correlation analysis and the feature importance comparisons support the predictive performance investigations within the studied context, as they offer some degree of interpretability and a better understanding of the technical problem under investigation. This latter important objective is also facilitated by the undertaken predictability comparison. The results are presented and discussed in detail with respect to their practical significance in Sections 3 and 4, respectively. Key recommendations for future research are also provided in Section 4, while the paper concludes with the study summary in Section 5.

## 2. Methods and data

### 2.1 Experimental dataset

We exploited information encompassed in the Catchment Attributes and MEteorology for Large-sample Studies (CAMELS) dataset, which is available in Newman et al. (2014) and Addor et al. (2017a). This dataset comprises a variety of catchment attributes, as well as minimum and maximum temperature, precipitation and streamflow time series at the daily time scale, from small- to medium-sized catchments spanning across the contiguous United States (Newman et al. 2015; Addor et al. 2017b), with the temperature and precipitation time series having been obtained by processing data by Thornton et al. (2014). From the entire dataset, we selected 511 geographical locations (see Figure 1), for which complete daily time series are available for the 34-year period between 1980 and 2013. For these specific geographical locations, we averaged the available daily minimum and maximum temperature time series values to compute 34-year-long time series of daily temperature means. These new temperature time series are hereinafter referred to simply as "daily temperature time series", as they were the ones analysed for feature extraction, together with the originally available daily precipitation and streamflow time series, as described in detail in Section 2.2.

Moreover, we selected the following catchment attributes with continuous values:
o   Logarithm of the mean elevation of the catchment (`log_elev_mean`);

- Logarithm of the mean slope of the catchment (`log_slope_mean`);
- Logarithm of the GAGESII estimate of the catchment area (`log_area_gages2`);
- Forest fraction of the catchment (`frac_forest`);
- Maximum monthly mean of the leaf area index of the catchment (`lai_max`);
- Green vegetation fraction difference of the catchment (`gvf_diff`);
- Dominant land cover fraction of the catchment (`dom_land_cover_frac`);
- Depth to bedrock of the catchment (`soil_depth_pelletier`);
- Soil depth of the catchment (`soil_depth_statsgo`);
- Maximum water content of the soil of the catchment (`max_water_content`);
- Sand fraction of the soil of the catchment (`sand_frac`);
- Silt fraction of the soil of the catchment (`silt_frac`);
- Clay fraction of the soil of the catchment (`clay_frac`);
- Water fraction of the soil of the catchment (`water_frac`);
- Organic material fraction of the soil of the catchment (`organic_frac`);
- Fraction of soil of the catchment marked as other (`other_frac`);
- Carbonate sedimentary rock fraction of the catchment (`carbonate_rocks_frac`);
- Subsurface porosity of the catchment (`geol_porosity`);
- Subsurface permeability of the catchment (`geol_permeability`).

The above-listed catchment attributes include three topographic, four land cover, nine soil and three geologic ones (reported in the same order from the top to the bottom), and are hereinafter referred to as "static catchment features" to be distinguished from the dynamic catchment features (i.e., the features that were obtained through time series analysis; see Section 2.2).

## 2.2 Time series analysis

We separately characterized each daily temperature, precipitation and streamflow time series (see Section 2.1) by computing its following features:

- Lag-1 sample autocorrelation of the time series (`x_acf1`);
- Sum of the squared sample autocorrelation values of the time series at the first ten lags (`x_acf10`);

- Lag-1 sample autocorrelation of the first-order differenced time series (`diff1_acf1`);
- Sum of the squared sample autocorrelation values of the first-order differenced time series at the first ten lags (`diff1_acf10`);
- Lag-1 sample autocorrelation of the second-order differenced time series (`diff2_acf1`);
- Sum of the squared sample autocorrelation values of the second-order differenced time series at the first ten lags (`diff2_acf10`);
- Lag-365 sample autocorrelation of the time series (`seas_acf1`);
- Lag at which the first zero crossing of the autocorrelation function is attained (`firstzero_ac`);
- Sum of the squared sample partial autocorrelation values of the time series at the first five lags (`x_pacf5`);
- Sum of the squared sample partial autocorrelation values for the first five lags of the first-order differenced time series (`diff1x_pacf5`);
- Sum of the squared sample partial autocorrelation values for the first five lags of the second-order differenced time series (`diff2x_pacf5`);
- Lag-365 sample partial autocorrelation (`seas_pacf`);
- Standard deviation of the first-order differenced time series (`std1st_der`);
- Number of times that the time series crosses the median (`crossing_points`);
- Spectral entropy of the time series (`entropy`);
- Number of flat spots in the time series (`flat_spots`);
- Lumpiness of the time series (`lumpiness`);
- Stability of the time series (`stability`);
- Nonlinearity of the time series (`nonlinearity`);
- Trend strength of the time series (`trend`);
- Strength of spikes in the time series (`spike`);
- Linearity of the time series (`linearity`);
- Curvature of the time series (`curvature`);

- Lag-1 sample autocorrelation of the remainder component of the time series, which is obtained after removing the trend and seasonal components (`e_acf1`);
- Sum of the squared sample autocorrelation values of the remainder component of the time series at the first ten time lags (`e_acf10`);
- Seasonality strength of the time series (`seasonal_strength`);
- Strength of peaks in the seasonal component of the time series (`peak`);
- Strength of troughs in the seasonal component of the time series (`trough`).

The above-listed features were selected because of their pronounced relevance to the field of stochastic hydrology and because of their interpretability. These features were computed after the time series were scaled to mean 0 and standard deviation 1. Moreover, for the computation of the features `diff1_acf1`, `diff1_acf10`, `diff2_acf1`, `diff2_acf10`, `diff1x_pacf5`, `diff2x_pacf5` and `std1st_der`, time series differencing was performed according to Hyndman and Athanasopoulos (2021, Chapter 9.1). For the computation of the features `trend`, `spike`, `linearity`, `curvature`, `e_acf1`, `e_acf10`, `seasonal_strength`, `peak` and `trough`, seasonal and trend decomposition using Loess (STL decomposition) was performed according to Hyndman and Khandakar (2008; see also Hyndman and Athanasopoulos 2021, Chapter 3.6), and by assuming 365 seasons per year. Additional information for (several of) the above-listed features can be found in data science works (e.g., Wang et al. 2006; Hyndman et al. 2015; Kang et al. 2017, 2020; Hyndman et al. 2020) and in books specialized on time series analysis (e.g., Box and Jenkins 1970; Wei 2006). This information is herein omitted for reasons of brevity, given also that the boxplot (and violin plot) summaries of the feature values over the 511 investigated catchments can support the perception of the above-provided definitions. These summaries are presented in Appendix B. Hereinafter, the various daily temperature, precipitation and streamflow features are alternatively referred to under the collective term "dynamic catchment features" to be distinguished from the static catchment features (see Section 2.1).

## 2.3 Correlation analysis

We investigated the relationships between potential predictor and dependent variables within regression-based streamflow regionalization contexts by computing Spearman correlations (Spearman 1904). Recall here that the static, daily temperature and daily precipitation features (see Sections 2.1 and 2.2) were investigated as predictor variables throughout this work, while the daily streamflow features (see Section 2.2) were investigated as dependent variables (see Figure 1). The results of the correlation analysis are presented in Section 3.1.

## 2.4 Feature importance comparisons

We applied explainable machine learning for comparing the static, daily temperature and daily precipitation features (see Sections 2.1 and 2.2) with respect to their relevance as potential predictors within the studied context for the daily streamflow features (see Section 2.2). More precisely, we studied 28 regression settings, each devoted to the prediction of a different daily streamflow feature. At each regression setting, we fitted random forests by Breiman (2001) with 2 000 trees and determined feature importance scores with permutation. This was made by following the unnormalized version of the implementation by Wright and Ziegler (2017). In brief, the applied procedure progresses as follows: For each tree, the mean square error is computed on the out-of-bag data. The same error metric is also computed after permuting each predictor variable. Then, the difference between the two outcomes is averaged over all trees. Once the permutation importance scores were obtained for a regression setting, the potential predictors were ranked based on them. Popularized information on the properties of random forests and their role in water science and informatics can be found in the review by Tyralis et al. (2019a), while discussions on the role of explainable machine learning in natural sciences can be found in the review by Roscher et al. (2020). The results of the feature importance comparisons are presented in Section 3.2.

## 2.5 Predictive performance comparisons

To further compare the proposed predictor variables with respect to their relevance within the studied context, we investigated the performance of random forests with 2 000 trees in predicting the various daily streamflow features (see Section 2.2) using seven different groups of predictor variables. These groups are the following ones (see Sections 2.1 and 2.2): {static catchment features}, {daily temperature features}, {daily precipitation features}, {static catchment features, daily temperature features}, {static catchment features, daily precipitation features}, {daily temperature features, daily precipitation features} and {static catchment features, daily temperature features, daily precipitation features}. The respective investigations were conducted under a *k*-fold cross validation setting, with *k* being equal to 10. Under this specific setting, the catchments were grouped into ten groups of approximately equal size. Ten different experiments were then conducted for each set {daily streamflow feature, group of predictor variables}, each time leaving out of the training process a different group of catchments, whose target feature values were subsequently predicted using the trained model and given the values of the predictors (with the target feature considered unknown).

Once the ten different experiments were finalized for a specific set {daily streamflow feature, group of predictor variables}, the root mean square error (RMSE) of the obtained predictions was computed. This computation was made collectively for all the 511 catchments; therefore, 28 (number of daily streamflow features) × 7 (number of groups of predictors) = 196 scores were obtained. Subsequently, and separately for each daily streamflow feature, the seven groups of predictors were ranked from the best (1st) to the worst (7th) based on their corresponding obtained scores. To assess the degree in which the predictive performance can differ when predicting each daily streamflow feature depending on the considered group of predictor variables, relative scores (taking the form of relative improvements in terms of RMSE) were also computed for the seven groups of predictors with respect to the group consisted only by the static catchment features. The results of the predictive performance comparisons are presented in Section 3.3.

## 2.6  Feature predictability comparison

To compare the various daily streamflow features (see Section 2.2) with respect to their predictability within regionalization contexts, we conducted scatterplots of their herein predicted versus observed values for the 511 investigated catchments. The predictions on these scatterplots are those obtained using the total of the static, daily temperature and daily precipitation features (see Sections 2.1 and 2.2) as predictor variables under a 10-fold cross validation setting (see Section 2.5). The results of the feature predictability comparison are presented in Section 3.4.

## 3.  Results

## 3.1  Feature correlations

Figure 2 presents the Spearman correlations between potential predictor and dependent variables within streamflow regionalization contexts. Overall, notable relationships exist that could be exploited for the regionalization of the studied daily streamflow features. As it is possibly expected, the relationships between daily precipitation and daily streamflow features are mostly more intense than the relationships between daily temperature and daily streamflow features. Still, the stability and trend strength of the daily temperature time series are also found to be notably related to several daily streamflow features. The same holds, to smaller extent, for some few other daily temperature features. Another observation regards the relationships between daily precipitation and daily streamflow features. Several daily streamflow features (e.g., `x_acf1`, `x_acf10`, `std1st_der`, `crossing_points`, `entropy`, `curvature`, `e_acf1`, `e_acf10`) exhibit approximately equally strong relationships with most of the daily precipitation features. Yet, there are also daily streamflow features that are not considerably related with daily precipitation features (e.g., `seas_pacf`, `stability`, `nonlinearity`, `peak`, `trough`), as well as others that are either considerably related only with few daily precipitation features (`flat_spots`, `lumpiness`, `linearity`) or related more to specific daily precipitation features than to others (`seas_acf1`, `spike`, `seasonal_strength`). Among the most considerable relationships revealed by our experiments, are also several ones between static catchment features and daily streamflow features. For instance, the mean elevation and the mean slope of the catchment, as well as the sand and clay fractions of its soil, are notably related to many daily streamflow features, including autocorrelation (e.g., `x_acf1`, `x_acf10`, `seas_acf1`), seasonality (`seas_acf1`, `seasonal_strength`) and others (e.g., `entropy`, `spike`).

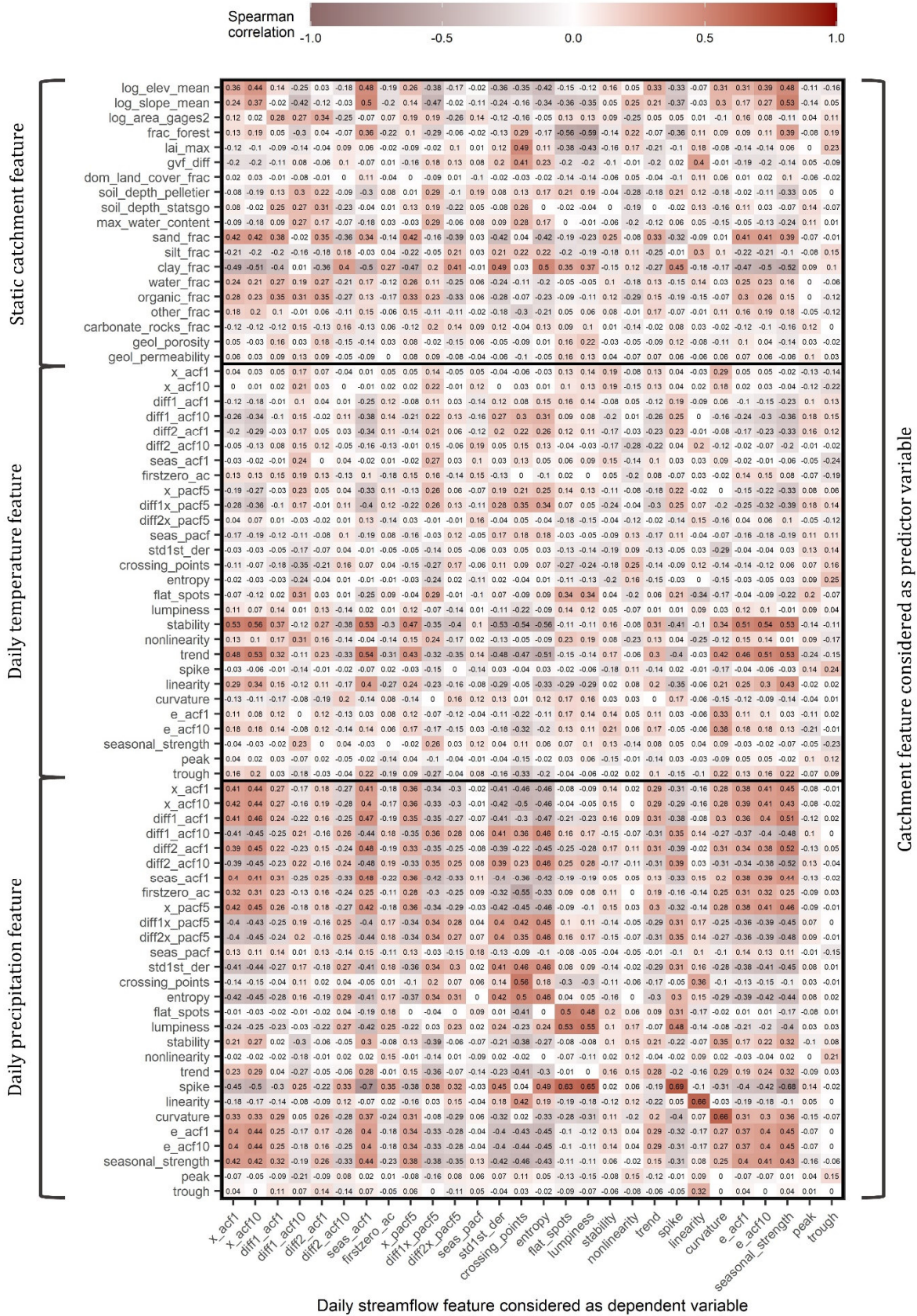

Figure 2. Spearman correlations between potential predictor and dependent variables of the investigated technical problem. The displayed feature abbreviations are explained in Sections 2.1 and 2.2.

## 3.2 Feature importance comparisons

Figure 3 presents the rankings of the static, daily temperature and daily precipitation features according to their usefulness in regionalizing the daily streamflow features. First, we observe that the most useful predictor variables, for most of the daily streamflow features, include catchment features from all the three investigated categories. We also observe that the most (least) relevant daily temperature characteristics within daily streamflow regionalization contexts are not the same with the most (least) relevant daily precipitation characteristics. For example, the features `stability` and `trend` of the daily temperature time series stand out for their relevance in predicting several daily streamflow features, while the features of the daily precipitation time series standing out for the same relevance are the following ones: `x_acf1`, `x_acf10`, `std1st_der`, `crossing_points` and `entropy`. This latter list is not exhaustive, with more daily precipitation features (other than `stability` and `trend`) being among the most important ones for generalizing streamflow features. Overall, Figure 2 can be used for interpreting Figure 3; however, this interpretation cannot be complete. Indeed, there are relationships of similar magnitudes that correspond to quite distant rankings (see, e.g., the Spearman correlations computed for the daily precipitation features `x_acf1`, `x_acf10`, `diff1_acf1`, `diff1_acf10`, `diff2_acf1`, `diff2_acf10`, `seas_acf1`, `firstzero_ac`, `x_pacf5`, `diff1x_pacf5` and `diff2x_pacf5` in comparison to their corresponding rankings), probably due to multicollinearity.

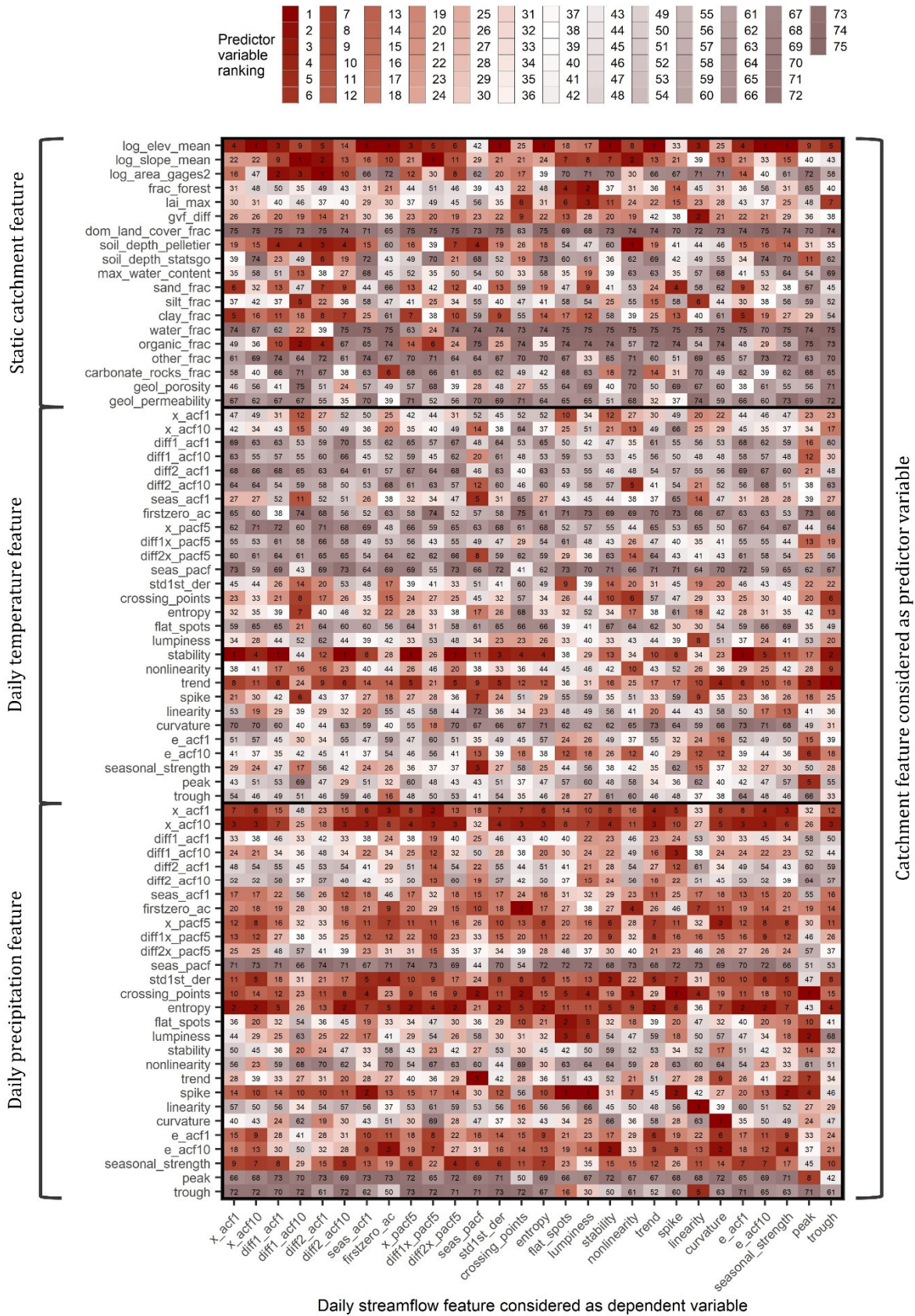

Figure 3. Rankings of potential predictor variables according to their importance in predicting various daily streamflow features. The displayed feature abbreviations are explained in Sections 2.1 and 2.2.

## 3.3 Predictive performance comparisons

Figure 4 presents the comparisons of the predictive performance in terms of RMSE of random forests, when the latter are used with seven different groups of predictors for the regionalization of various daily streamflow features. First, we observe that, for half of the daily streamflow features, the use of the total of the static, daily temperature and daily precipitation features as predictor variables offers the smallest RMSE (see Figure 4a). We also observe that, for the remaining daily streamflow features, the relative scores corresponding to the utilization of (a) the total of the static, daily temperature and daily precipitation features as predictors and (b) the best group of predictors differ less than 4% (see Figure 4b). Furthermore, we observe that, while the static features constitute a sufficient group of predictors alone for some daily streamflow features, they are less informative than daily temperature and/or precipitation features for predicting other daily streamflow features (see Figure 4a). Notably, the utilization of the daily temperature and/or precipitation features as predictors in our experiments, especially when static catchment features were also utilized, lead to considerably large predictive performance improvements for several daily streamflow features (`seas_acf1`, `flat_spots`, `lumpiness`, `spike`, `linearity`, `curvature`, `seasonal_strength`, `peak`, `trough`). These improvements were found up to approximately 17% (see Figure 4b).

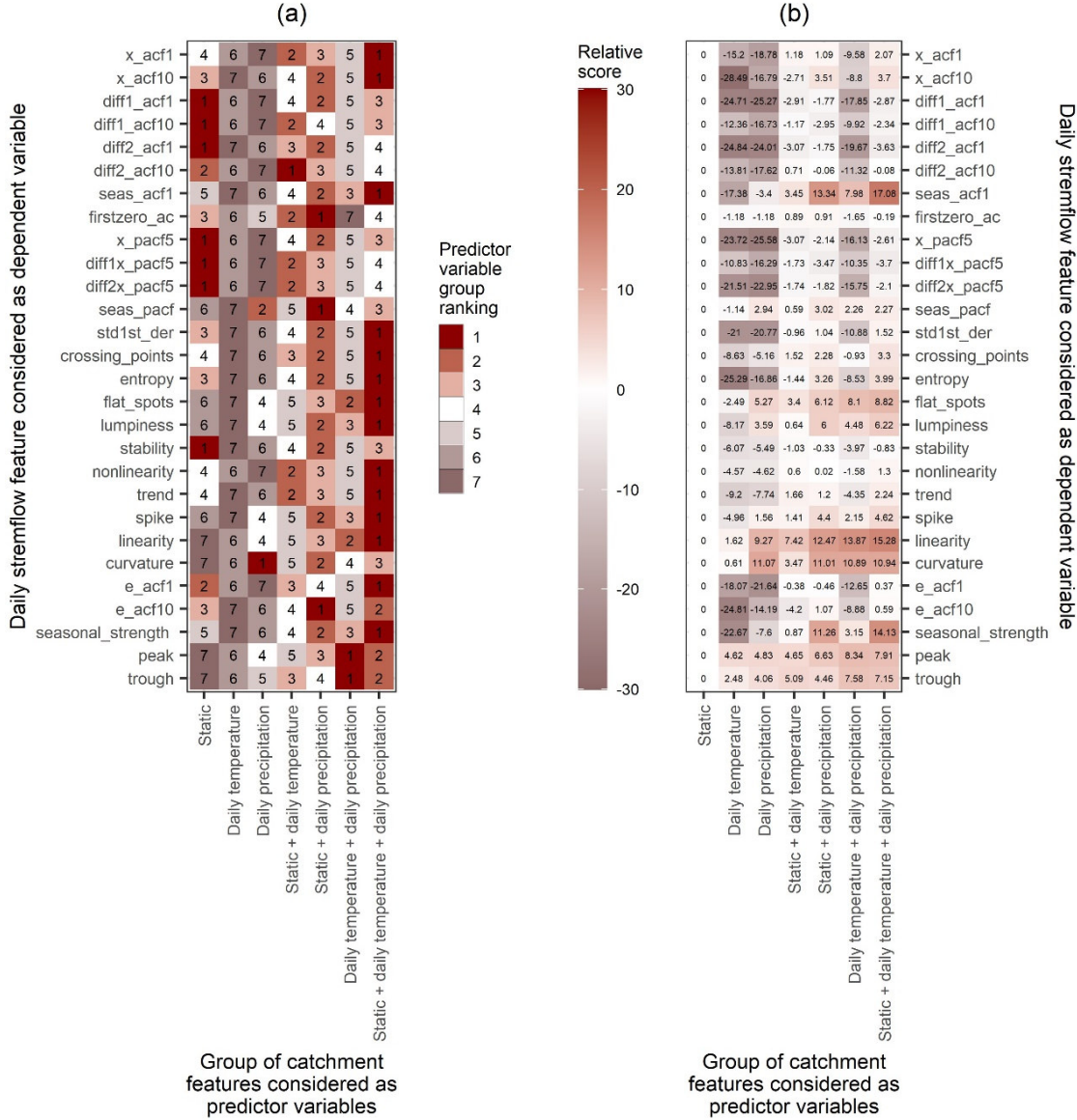

Figure 4. Performance comparisons between seven prediction methods that use random forest with different catchment feature groups as predictor variables. These comparisons were made over the 511 investigated catchments under a 10-fold cross validation setting. They take the form of (a) rankings of the predictor variable groups from the best (1st) to the worst (7th) performing ones in terms of root mean square error, and (b) relative performance differences (%) with respect to using only the static features as predictor variables. The displayed feature abbreviations are explained in Section 2.2.

### 3.4 Feature predictability comparison

Figures 5 and 6 present examples of predicted versus observed daily streamflow features, thereby allowing us to (roughly) assess and compare the daily streamflow features with respect to their predictability within regionalization contexts. Indeed, this predictability strongly depends on the daily streamflow feature. Examples of daily streamflow features that are notably more difficult than others to regionalize include the following ones:

`seas_pacf`, `nonlinearity`, `peak` and `trough` (see Figure 5l, s and Figure 6g, h). Note here that these four features are also among the least related ones to the static, daily temperature and daily precipitation features (see Figure 2). On the other hand, there are features that stand out because of their high (higher) predictability with regionalization contexts. Such features include, among others, the following ones: `x_acf1`, `seas_acf1`, `std1st_der`, `entropy`, `e_acf1` and `seasonal_strength` (see Figure 5a, g, m, o and Figure 6d, f).

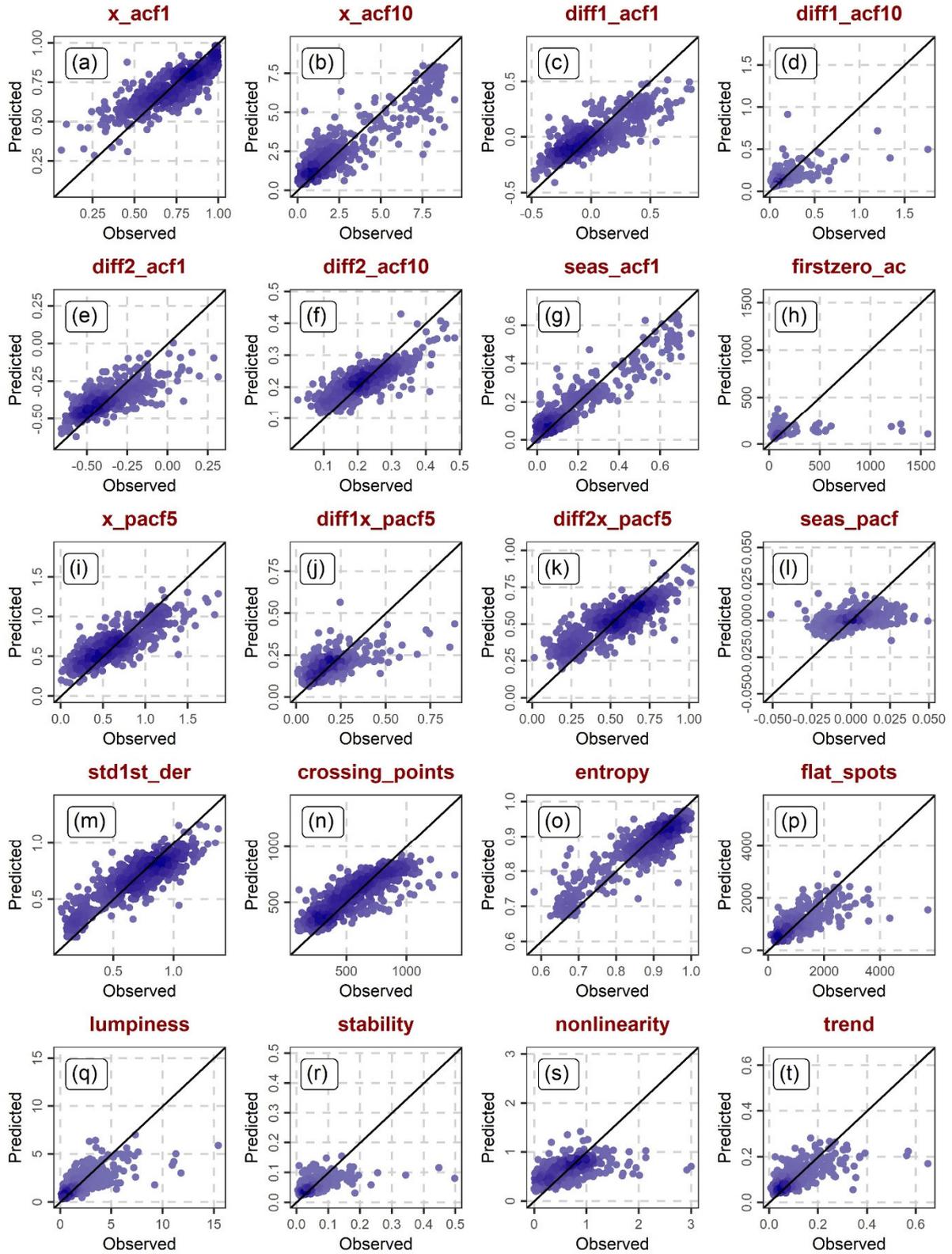

Figure 5. Predicted versus observed daily streamflow features for the 511 investigated catchments when using as predictor variables the total of the static, daily temperature and daily precipitation features under a 10-fold cross validation setting (part 1). The displayed feature abbreviations are explained in Section 2.2.

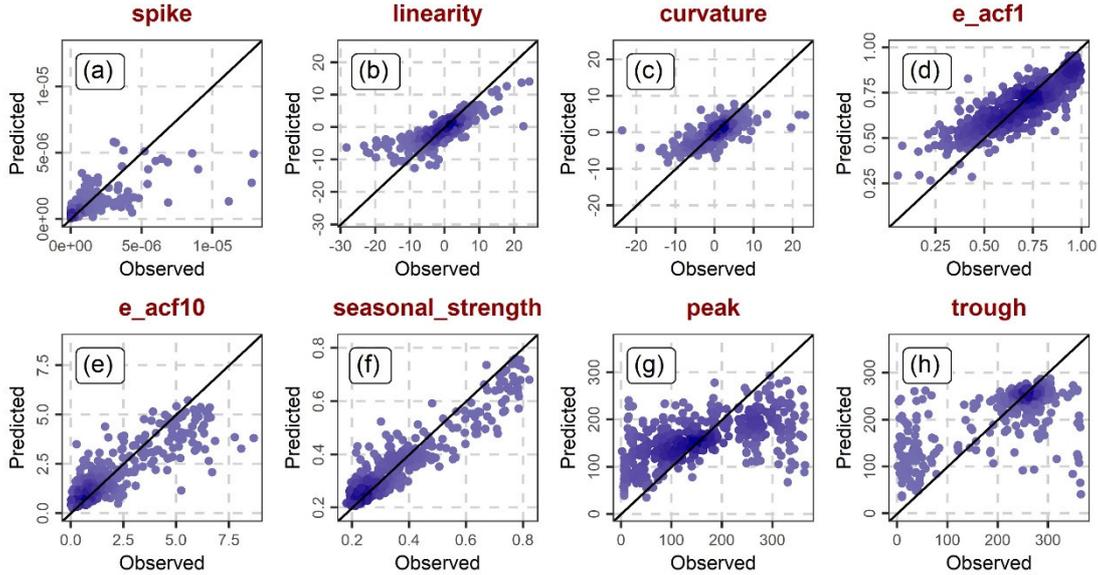

Figure 6. Predicted versus observed daily streamflow features for the 511 investigated catchments when using the total of the static, daily temperature and daily precipitation features as predictor variables under a 10-fold cross validation setting (part 2). The displayed feature abbreviations are explained in Section 2.2.

## 4. Discussion

Our contribution to the literature consists in the introduction of new methodological elements into the area of regression-based streamflow regionalization, as well as in the successful and extensive utilization of these elements within a detailed framework that aimed to deliver useful insights into (a) the various relationships exploited for the regionalization and (b) the predictability of the streamflow features in the relevant regression settings. Overall, the predictions issued in our experiments were sufficient, given also that their target values do not depend on the magnitude of the time series, in contrast to several hydrological signatures from the catchment hydrology field (see, e.g., part of the ones investigated by Addor et al. 2018; Tyralis et al. 2021b). Indeed, it might be more difficult to regionalize streamflow autocorrelation features (and the remaining features examined herein) than it is to regionalize mean annual streamflow (see, e.g., the related notes on the uncertainty in regionalization results by Westerberg et al. 2016). Of course, such comparisons are not covered by our aims and experiments, and could be the subject of future investigation and discussion.

To our view, one of the most notable findings of this work is that both static and dynamic catchment features are considerably relevant to regionalizing a variety of daily streamflow features, including autocorrelation, entropy, temporal variation, seasonality and other daily streamflow features. More precisely, among the most useful predictors are the mean elevation of the catchment, the spectral entropy, the seasonality strength and autocorrelation features of the daily precipitation time series, as well as the stability and trend strength of the daily temperature time series. Further than this, we also identified daily streamflow features that are more (or less) difficult to regionalize than others, similarly to what it is regularly made for hydrological signatures in catchment hydrology (see, e.g., the work by Addor et al. 2018 and its relevant discussions). Interestingly, among the most regionalizable daily streamflow features are the spectral entropy, the seasonality strength and several autocorrelation ones, which are also notably relevant within feature-based time series simulation frameworks (such as the one proposed by Kang et al. 2020) from a conceptual point of view. Given this relevance, the methodological elements introduced in this work could contribute substantially to the reduction of modelling and simulation uncertainties. A further reduction through the regionalization of features that are less difficult to regionalize (and through a possible subsequent formation of feature-based time series simulation frameworks around them and around more regionalizable features) could also be considerable and is, thus, worthy of investigation.

Of course, the above-discussed findings of this work concern the daily time scale only and could differ for other time scales in ways that could be investigated in the future. The various open research pathways further include the simultaneous consideration of the spatial proximity of the stations and time series features for obtaining predictions, as well as the utilization of additional time series features from the stochastic (statistical) hydrology and the data science fields, and even the utilization of the concept of massive feature extraction (Fulcher et al. 2013; Fulcher and Jones 2014, 2017; Fulcher 2018) from the latter of the aforementioned fields. In fact, although this work already covers a larger variety and a larger number of time series features than usual in stochastic hydrology, many more time series features are available (see, e.g., the ones investigated by Hamed 2008; Montanari 2012; Ledvinka 2015; Ledvinka and Lamacova 2015; Juez and Nadal-Romero 2020, 2021; Papacharalampous et al. 2021, 2022) and could be useful in streamflow regionalization settings, given also the generally acknowledged significance of finding new informative predictors for obtaining improved predictions.

Another open research endeavour rotates around the regression algorithm of the regionalization, because of its own core importance for improving streamflow feature predictions in large-sample hydrology. The experiments conducted herein were solely based on random forests. Nonetheless, more regression algorithms (see, e.g., those listed and documented by Hastie et al. 2009; James et al. 2013, as well as those investigated in other contexts in hydrometeorology by Tyralis et al. 2021a; Zhang and Ye 2021) are worthy of investigation together with a variety of time series features in streamflow regionalization contexts, with boosting being an appealing option among them because of its theoretical properties (see, e.g., Tyralis and Papacharalampous 2021a, Section 3) and its relevance to determining feature importance within explainable machine learning settings. For these same reasons, boosting algorithms were previously proposed and extensively investigated by Tyralis et al. (2021b) for predicting various hydrological signatures probabilistically. Also notably, methodological elements of this latter work could be borrowed for extending the herein proposed methodological framework for the probabilistic prediction of streamflow features within regionalization settings. Other machine and statistical learning algorithms that could be exploited in a straightforward manner in this regard can be found in Papacharalampous et al. (2019; see also the references therein), and fall into the larger category of quantile regression algorithms. Most of these algorithms base their training on the quantile loss (else referred to as the "pinball loss") function, the utilization of which was also proposed by Tyralis and Papacharalampous (2021b) for converting (even more) interpretable models into probabilistic ones.

## 5.   Summary and conclusions

In this work, we proposed new concepts and methodological elements for supporting the transfer from gauged to ungauged locations of information that is useful for streamflow description and modelling, driven by central themes appearing in stochastic (statistical) hydrology and, at the same time, driven by the observation that these specific themes were scarce in (if not absent from) the previous literature on streamflow regionalization. More precisely, we proposed the estimation of a large variety of time series features (including autocorrelation, partial autocorrelation, entropy, temporal variation, seasonality, trend, lumpiness, stability, nonlinearity, linearity, spikiness, curvature and other features) from large multi-site datasets comprising temperature, precipitation and streamflow information, and the subsequent transfer of streamflow feature information from gauged to ungauged sites by considering the various temperature and precipitation features, together with other (e.g., traditional) catchment attributes, as predictor variables within regression-based frameworks.

The relevance of the proposed streamflow regionalization strategy was illustrated through extensive large-sample investigations, which were conducted for 511 small- to medium-sized catchments spanning across the contiguous United States, and involved the estimation of 28 time series features for 34-year-long daily temperature, precipitation and streamflow time series originating from these catchments. Once this estimation was complete, the temperature, precipitation and streamflow features were merged with traditional topographic, land cover, soil and geologic attributes within regression-based streamflow regionalization frameworks. In this context, we found the mean elevation of the catchment, the spectral entropy, the seasonality strength and several autocorrelation features of the precipitation time series, and the stability and trend strength of the temperature time series, to be among the most useful predictors for many streamflow features, while we additionally provided a possible (rough) interpretation of these specific findings by examining the relationships between the various potential predictor and independent variables in terms of correlations. Lastly, we found the spectral entropy, the seasonality strength and several autocorrelation features of the streamflow time series to be more regionalizable than others.

**Author contributions:** GP and HT contributed equally to all the aspects of this work.

**Acknowledgments:** The authors are sincerely grateful to the Journal for inviting the submission of this paper. They also wish to acknowledge constructive remarks by the Editor and Reviewers.

## Appendix A   Statistical software information

The computations and visualizations were performed in R Programming Language (R Core Team 2021). The following contributed R packages were utilized: `caret` (Kuhn 2021), `cowplot` (Wilke 2020), `data.table` (Dowle and Srinivasan 2021), `devtools` (Wickham et al. 2021), `gdata` (Warnes et al. 2017), `gridExtra` (Auguie 2017), `hydroGOF` (Zambrano-Bigiarini 2020), `knitr` (Xie 2014, 2015, 2021), `MASS` (Venables and Ripley 2002; Ripley 2021), `ranger` (Wright and Ziegler 2017; Wright 2021), `rmarkdown` (Xie et al. 2018; Xie et al. 2020; Allaire et al. 2021), `stringi` (Gagolewski 2021), `tidyverse` (Wickham et al. 2019; Wickham 2021), `tsfeatures` (Hyndman et al. 2020).

## Appendix B  Dynamic catchment features

Summaries of the dynamic catchment features (see Section 2.2) over the 511 investigated catchments are provided in Figures B1 and B2.

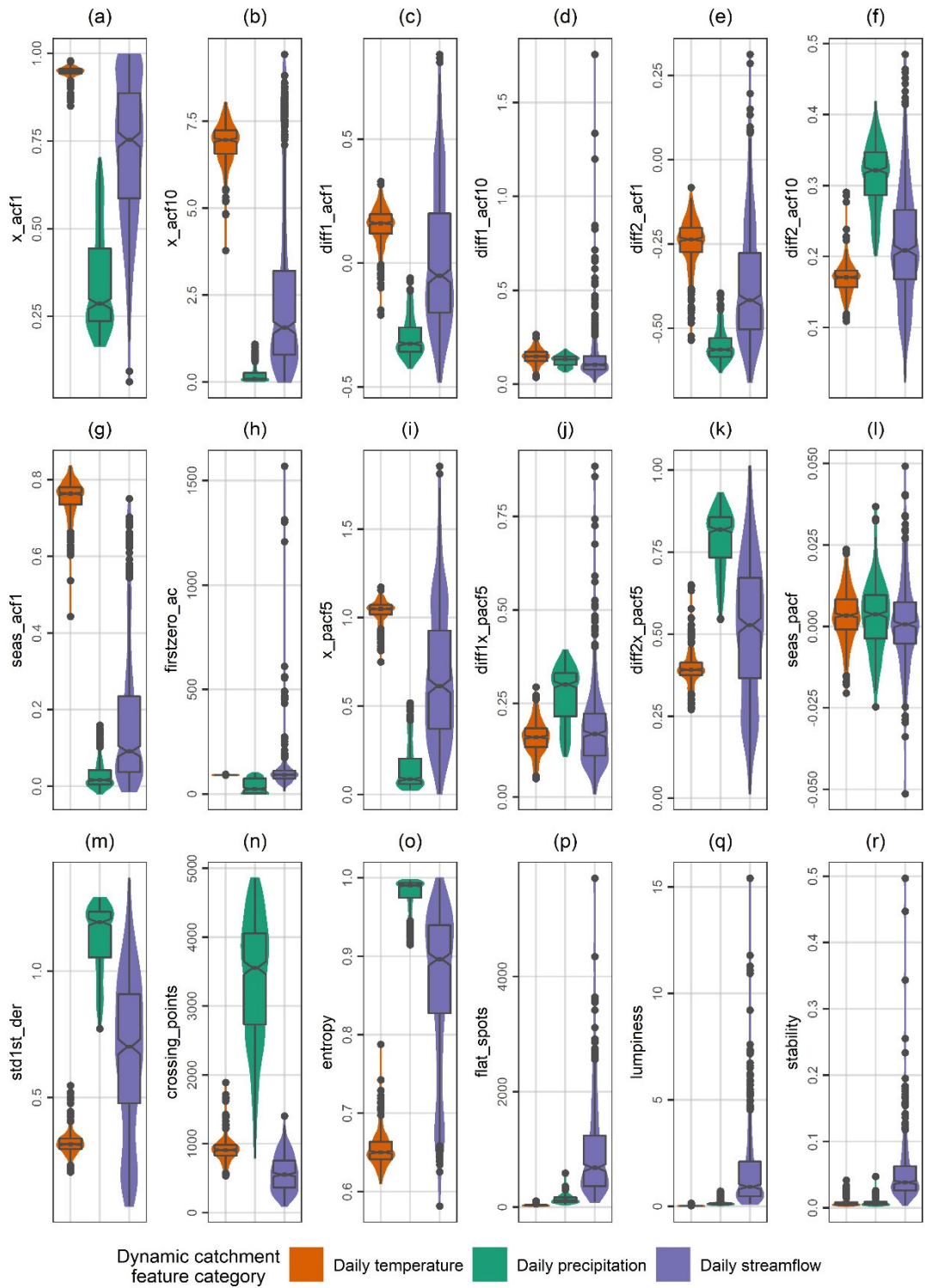

Figure B1. Summaries over the 511 investigated catchments of the estimated daily temperature, precipitation and streamflow features (part 1). For the explorations and predictions, these dynamic catchment features were merged with static catchment features. The displayed feature abbreviations are explained in Section 2.2.

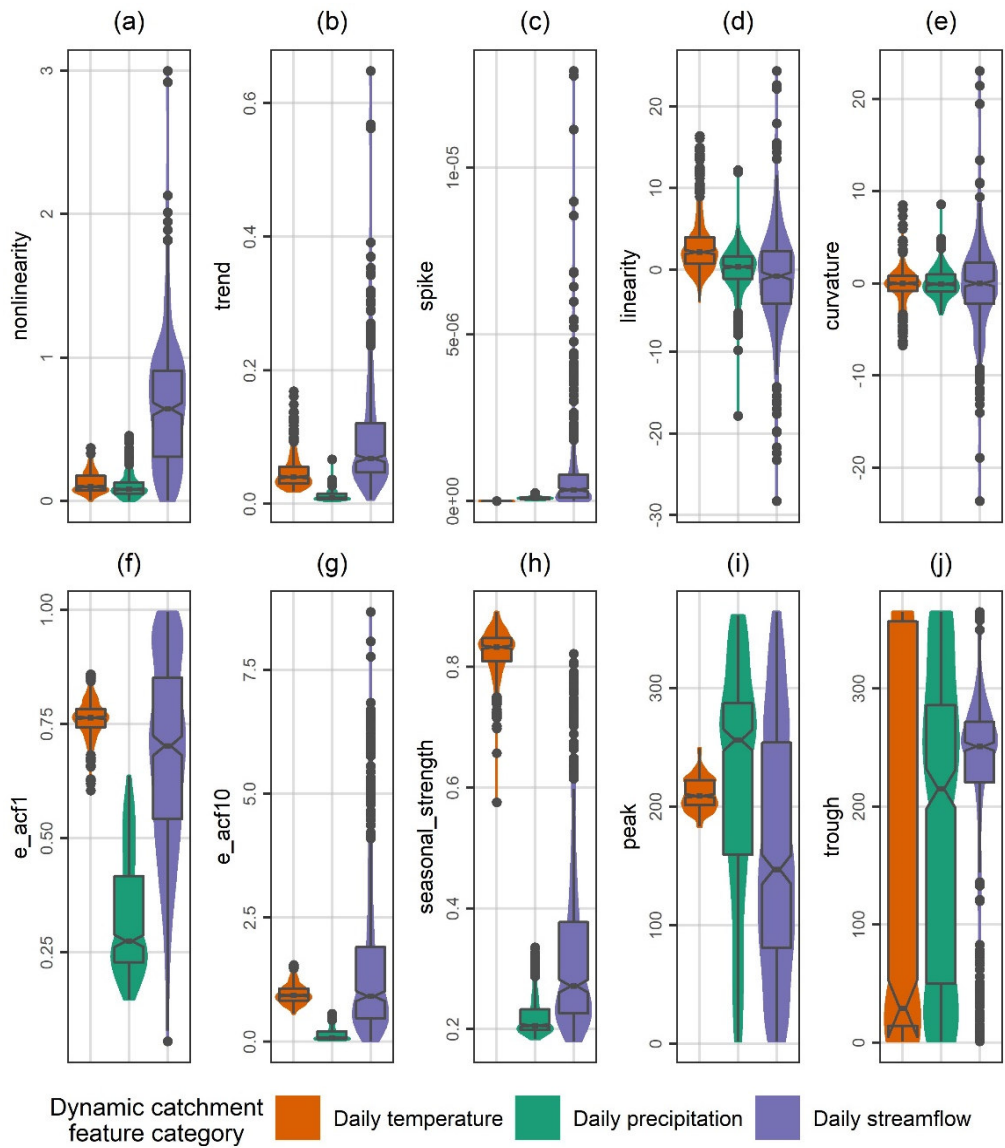

Figure B2. Summaries over the 511 investigated catchments of the estimated daily temperature, precipitation and streamflow features (part 2). For the explorations and predictions, these dynamic catchment features were merged with static catchment features. The displayed feature abbreviations are explained in Section 2.2.